\documentclass[twocolumn,showpacs,preprintnumbers,amsmath,amssymb]{revtex4}

\usepackage{graphicx}% Include figure files
\usepackage{dcolumn}% Align table columns on decimal point
\usepackage{bm}% bold math

\def\beqn{\begin{eqnarray}}
\def\eeqn{\end{eqnarray}}

\begin{document}

\preprint{hep-th/0702016
}

\title{Multi-Particle States in Deformed Special Relativity}

\author{S.~Hossenfelder}
 
 \email{sabine@perimeterinstitute.ca}
\affiliation{Perimeter Institute\\ 
31 Caroline St. N, Waterloo, Ontario, N2L 2Y5, Canada}%

\date{\today}% It is always \today, today,
             %  but any date may be explicitly specified

\begin{abstract}
We investigate the properties of multi-particle states in Deformed Special Relativity ({\sc DSR}).
Starting from the Lagrangian formalism with an energy dependent metric, the conserved Noether current 
can be derived which is additive in the usual way. The integrated Noether current had previously been 
discarded as a conserved quantity, because it was correctly realized that it does no longer obey 
the {\sc DSR} transformations. We identify the reason for this mismatch in the fact that {\sc DSR} depends
only on the extensive quantity of total four-momentum instead of the 
energy-momentum densities as would be appropriate for a field theory. We argue that the 
reason for the failure of {\sc DSR} to reproduce the standard transformation behavior in the 
well established limits is due to the missing sensitivity to the volume inside which 
energy is accumulated. We show that the soccer-ball problem is absent if one formulates {\sc DSR} instead for the
field densities. As a consequence, estimates for predicted effects  have to be corrected by 
many orders of magnitude. Further, we derive that the modified quantum field theory 
implies a locality bound. 

\end{abstract}

%PACS, the Physics and Astronomy
                             % Classification Scheme.
%\keywords{Suggested keywords}%Use showkeys class option if keyword
                              %display desired
\maketitle

\section{Introduction}

The phenomenology of quantum gravity has received increased attention during the last years.
In the absence of a testable theory quantum gravity, predictions based on effective models have
been studied which use only some few well motivated assumptions. One
such assumption is the presence of a regulator in the ultraviolet, or the existence of a
maximal energy scale respectively. The requirement that Lorentz-transformations in momentum
space have this scale as a second invariant leads 
to a class of Deformations of Special Relativity ({\sc DSR}).
As one of the most general expectations arising from a theory of quantum gravity, 
these modified Lorentz transformations have been studied extensively \cite{Amelino-Camelia:2000ge,Amelino-Camelia:2000zs,Jacobson:2001tu,Amelino-Camelia:2002vw,Sarkar:2002mg,Konopka:2002tt,Alfaro:2002ya,Heyman:2003hs,Jacobson:2003bn,Magueijo:2001cr,Magueijo:2002am,Ahluwalia-Khalilova:2004dc,Smolin:2005cz,Hossenfelder:2006rr}.

However, despite the fact that it is possible to use kinematic arguments to predict 
threshold corrections, a fully consistent quantum field theory with {\sc DSR} is still not available.
Though there are notable attempts \cite{Hossenfelder:2003jz,Hossenfelder:2006cw,Konopka:2006fh,Magueijo:2006qd,Ghosh:2006cb}, 
one has faced serious conceptual problems in the formulation of a field theory, such as
the proper definition of conserved quantities in interactions, and the transformation of multi-particle
states, also known as the soccer-ball problem.
 
In this paper, we argue that the reason for this mismatch lies in the investigation of extensive
quantities like total energy and momentum, rather then intensive quantities like energy- and momentum
densities which would be appropriate for a field theory. Originally, {\sc DSR} was formulated 
\cite{Amelino-Camelia:2000ge,Amelino-Camelia:2000zs} as a (classical)
theory for a point particle, and it has been shown \cite{Kowalski-Glikman:2002ft,Kowalski-Glikman:2003we} that {\sc DSR} can be understood
as a deformation of the momentum space that belongs to the point particle. However, {\sc DSR} -- through the very introduction of a minimal length --
implies a generalized uncertainty principle \cite{Hossenfelder:2005ed,Kempf:1994su,ml1,ml2,ml3}, which forbids it to 
localize a particle to a point. Therefore, already this formulation must be interpreted as a theory for an 
energy distribution with maximally possible localization, and the momentum space properties for 
space-time points inside this space-time volume.

If one wants to construct a field theory that consistently incorporates DSR, the transformation
behavior for a classical particle with four momentum {\bf p} can not independently be transferred to each of the single 
field's modes,  since superposition of these modes implies that the properties at a point in spacetime - and
therefore the momentum space at this point - depend not only on the single mode but on the energy density of all
the present excitations. 
 
Once one realizes that the quantity to be bounded by the Planck scale 
should not be the total energy of a system, but rather its energy density, the soccer-ball problem vanishes
and multi-particle states transform appropriately. Unfortunately, the energy density
of all experimentally accessible objects is far too small to make any quantum gravitational effects
of this kind important. Thus, if one  formulates the quantum field theory with {\sc DSR}, the so
far proposed predictions are unobservable. 

This paper is organized as follows. In the next section we introduce the notation. In section \ref{mult} we
briefly recall the problem of multi particle states and examine its cause. Section \ref{qft} summarizes the
quantum field theory formalism previously used in  \cite{Hossenfelder:2003jz,Hossenfelder:2006cw,Hossenfelder:2005ed}. In the following section \ref{soc}, we investigate the soccer-ball problem and show how it can be resolved. Predictions are
revisited in section \ref{pred}. The discussion and the conclusions can be found in section \ref{conc}. 

Throughout this paper we use the convention $\hbar=c=1$,
such that the Planck mass is the inverse of the Planck length $m_{\rm p} = 1/l_{\rm p}$. Bold faced quantities 
{\bf p}, {\bf q} are four-vectors. Capital Latin label particles. Small Greek indices, and small Latin indices from the beginning of the alphabet run from 0 to 3 and label space-time coordinates. Quantities with small Greek indices transform under standard Lorentz-transformation; quantities with small Latin indices from the beginning of the alphabet transform under
the deformed transformations. The indices $k$ and $p$ refer to the 
wave-vector  and momentum, respectively.

\section{Deformed Special Relativity}
\label{dsr}

With use of the notation introduced in \cite{Hossenfelder:2003jz,Hossenfelder:2006cw,Hossenfelder:2005ed}, 
the quantity ${\bf p} = (E,p)$ transforms as a standard Lorentz-vector, and is distinguished from 
${\bf k} = (\omega,k)$, which obeys the modified transformation that is non-linear in 
${\bf k}$. The former quantity {\bf p} can always be introduced, the important step is eventually its 
physical interpretation.  In the standard {\sc DSR} formalism, {\bf p} is often referred to as
the pseudo-momentum.

A general relation between {\bf p} and {\bf k} can be of the form
\beqn
{\bf k} = F({\bf p}) = \left( E~f({\bf p}), p~g({\bf p})\right)\quad, \label{defs}
\eeqn
with the inverse ${\bf p} = F^{-1}({\bf k})$, that we will denote for better readability as 
$F^{-1}({\bf k})\equiv G({\bf k})$. 
As examined in \cite{Hossenfelder:2005ed} 
these theories can, but need not necessarily have an energy dependent speed of light. 
An obvious requirement is that the function $F$ reduce to multiplication with $\hbar$ in the limit
of energies being small with respect to the Planck scale. In order to implement a maximum energy
scale, either one or all components of {\bf k} should be bounded by $m_{\rm p}$. In these theories, one has
a modified dispersion relation ({\sc MDR}) of the form $G({\bf k})^2 = m^2$.

It is now straight forward to derive the transformation that maps ${\bf k} \to {\bf k}'$ when applying a Lorentz-boost,
and which respects the invariance of the modified dispersion relation. One just keeps in mind that the relation 
for {\bf p} is the standard relation $p^2 - E^2 = p'^2 - E'^2$,
from which one finds the standard Lorentz transformation in momentum space. We will denote these standard transformations with ${\bf p}' =  L({\bf p})$. Then one gets the modified  Lorentz-transformations acting on {\bf k} by requiring
\beqn 
{\bf k}' = F({\bf p}') = F( L({\bf p})) = F(L(G({\bf k}))) \quad. \label{trafok}
\eeqn
We will denote these transformations as ${\bf k}' = \widetilde{L}({\bf k})$. The transformations Eqs. (\ref{trafok}) will be non-linear 
in $(\omega,k)$ since $F$ is. By construction, implemented upper bounds on one or all components of {\bf k}
are respected.  
For special choices of $F$ one finds the {\sc DSR} transformations used in the literature. An explicit
example \cite{Magueijo:2001cr} is $f(E) = g(E) = 1/(1+ E/m_{\rm p})$, for which one has the transformations
\beqn
\omega' &=& \frac{\gamma (\omega - v k)}{1 - \omega/m_{\rm p} + \gamma(\omega -v k)/m_{\rm p}}\quad, \nonumber\\
k' &=& \frac{\gamma ( k - \frac{v}{c^2} \omega)}{1 - \omega/ m_{\rm p} + \gamma (\omega -v k)/m_{\rm p}} \quad.
\eeqn

One can understand {\sc DSR} as a theory with a curved momentum space. In fact, if one integrates over all
possible values of {\bf k}, and rewrites the integration into momentum space one finds
\beqn
\int d^4 k = \int d^4 p \bigg| \frac{\partial F}{\partial p}\bigg| \label{volume}
\eeqn
where the quantity under the right integral is the Jacobian determinant. This can be read as a curved
momentum space with an energy dependend metric \cite{Magueijo:2002xx,Kimberly:2003hp,Hinterleitner:2004ny} $g$ and $|\partial F / \partial p| = \sqrt{-g}$. The most
extensively investigated geometry is that of DeSitter space \cite{Kowalski-Glikman:2002ft,Kowalski-Glikman:2003we}. 
For the cases investigated in \cite{Hossenfelder:2003jz,Hossenfelder:2006cw,Hossenfelder:2005ed}, the geometry is conformally
flat \footnote{For these cases, observables do not depend on the choice of coordinates in momentum space, and
the geometrical properties are the only relevant input.}.

However, one
should keep in mind that here we have considered single particles, and the momentum space we were referring
to was the momentum space of that particle, not a global property. In fact, if one considers
a field theory, every point of our space-time manifold should have a corresponding momentum space, and its
properties can in principle be a function of the space-time coordinates \footnote{Note that we are interested in
the actual space-time coordinates and not in the configuration space.}. In the limit where 
quantum gravitational effects are negligible, one would expect to a flat momentum space to be a very
good approximation, and to recover the standard transformation laws of Special Relativity.

\section{Multi-Particle States}
\label{mult}

So far we have considered only one particle. The question how to generalize the formalism
of {\sc DSR} to multi-particle states is essential if one wants to formulate a quantum field theory. The
missing description of multi-particle systems is an huge obstacle on
the way to formulate the principles of the theory, and to recover the limiting cases 
of the Standard Model and Special Relativity.
Though large progress has been made regarding the solution of this
problem \cite{Magueijo:2006qd,Hinterleitner:2004ny,Judes:2002bw,Girelli:2004ue,Girelli:2006ez}, the issue is still not completely settled
and open questions remain \cite{Ahluwalia-Khalilova:2004dc,Liberati:2004ju,Toller:2003tz}.

In particular,
one wants to construct a conserved quantity for bound states and interactions. Let us consider a
two particle system with ${\bf p}_A, {\bf p}_B$ or ${\bf k}_A, {\bf k}_B$ respectively, and ask for the conserved quantity ${\bf q}$. The most obvious choice is
\beqn
{\bf q} = {\bf p}_A + {\bf p}_B \label{c1}\quad,
\eeqn
which transforms as a usual Lorentz vector, and is the way pursued in 
\cite{Hossenfelder:2003jz,Hossenfelder:2006cw,Hossenfelder:2005ed}.  However, this option is admittedly not very exciting, and it has been pointed out \cite{Amelino-Camelia:2000mn} that in fact within {\sc DSR} the construction of a conserved quantity seems to be not uniquely defined. The next obvious choice that one would take is
\beqn
{\bf q} = {\bf k}_A + {\bf k}_B \label{c2}\quad.
\eeqn
However, if one requires ${\bf q}$ to obey the same deformed transformations as ${\bf k}_A$ and  ${\bf k}_B$,  
then this quantity does not transform properly. Since the transformations $\widetilde{L}$ are not linear, one has
\beqn
{\bf q}' &=& \widetilde L({\bf q}) = \widetilde L({\bf k}_A + {\bf k}_B) \nonumber \\
&\neq& \widetilde L({\bf k}_A) + \widetilde L ({\bf k}_B) = {\bf k}'_A + {\bf k}'_B \quad.  \label{notra}
\eeqn
Note that this arises from the fact that the transformation acting on the ${\bf k}_A$ (${\bf k}_B$)  is a function of ${\bf k}_A$ (${\bf k}_B$)  only
and not of the total conserved quantity. Since the transformation behavior reflects the properties
of the curved momentum space of the particle with energy ${\bf k}$, this means that these momentum spaces are
independent of each other. This is a justified expectation if the particle's energies are sufficiently localized, as
not to influence each other. This is appropriate for point particles, but 
will definitely not be the case for plane waves. 

One could of course just define the transformation behavior of ${\bf q}$ to
be equal to that of ${\bf k}_A + {\bf k}_B$. But then, the transformation of the conserved charge ${\bf q}$ would 
depend on the decomposition into the added quantities and not be unique. I.e. another decomposition into 
${\bf q} = {\bf k}_C + {\bf k}_D$ would lead to a different transformation behavior. Though this seems 
unintuitive, and we will not further examine this transformation law, we would like to point out that this 
possibility remains an option.
 
When one discards the addition law (\ref{c2}), one is then lead to the conclusion that 
the quantity ${\bf k}$ has to obey a modified addition law,  which we will denote with $\oplus$, and   which is given by
\beqn
{\bf k}_A \oplus {\bf k}_B = F({\bf p}_A + {\bf p}_B) \label{c3}\quad.
\eeqn
In such a way, one can define
\beqn
{\bf q} = {\bf k}_A \oplus {\bf k}_B   \quad,
\eeqn
which transforms appropriately under applying the transformation (\ref{trafok})
\beqn
{\bf q}' &=& F( L ({\bf p}_A + {\bf p}_B )) = F( {\bf p}'_A + {\bf p}'_B ) \quad.
\eeqn 
If one considers an interaction of the type $A + B \to C$, and identifies the energy of the particle
$c$ with the above defined quantity ${\bf q}$ one obtains the conservation law
\beqn
0 = {\bf q} - {\bf k}_A \oplus {\bf k}_B \quad.
\eeqn
This conservation law deviates from the standard prescription due to the modified addition law. This
gives rise to the predicted threshold modifications. In case one considers more than three particles, one
has more choices for the non-linear addition \cite{Amelino-Camelia:2000mn}. Note that ${\bf k}_A$ is an
element of particle $A$'s phase space, whereas ${\bf k}_B$ belongs to particle $B$'s space. The addition
therefore is not performed inside the single particle phase-space, but instead defines a structure on the
multi particle phase-space.  

However, with the prescription Eq. (\ref{c3}) one runs into another problem. By construction, the function $F$ 
creates an upper bound on ${\bf k}$. Unfortunately, we know that bound systems of elementary particles can
very well exceed the Planck mass, and this {\sc DSR} formalism therefore can not apply for them. 
The reason for this mismatch, also known as the soccer-ball problem, is the non-linearity of the 
transformations, which should be suppressed when the number of constituents grows.  

One should also note that the addition law has been chosen and not been derived, which means
it is an additional assumption of {\sc DSR}.

\section{Towards a Field Theory}
\label{qft}

One way or the other, if {\sc DSR} is a well defined symmetry principle, it should be possible to just derive
the conserved quantity for multi-particle states, and resolve the soccer-ball problem. Indeed, this is
straight-forward to do, as has been shown e.g. in \cite{Hossenfelder:2006cw}. 

In the following we will use the formalism with an energy dependent metric $g_{\mu \nu}({\bf k})$  that has been introduced
and worked out in \cite{Kimberly:2003hp,Hossenfelder:2006cw}. The momentum is denoted by $p^i$ and transforms under the usual Lorentz-transformation. The wave-vector is obtained by converting the index with an energy dependent field that we will denote with $h$. Since the relation between both momentum and wave-vector depends on the energy, the transformation of the wave-vector will no longer be the standard Lorentz-transformation. One also notices that the volume element in momentum space becomes energy dependent as previously mentioned (compare to Eq. (\ref{volume})). 

Under quantization, the metric becomes an operator $g_{\mu \nu} (\partial)$. The energy dependence of the metric can be interpreted as a backreaction effect on the propagating particle: 
If the energy density in a space-time region reaches the Planckian regime, then the particle will significantly disturb the
background it propagates in. In the limit where the metric approaches that of flat Minkowski space one recovers standard
Special Relativity.

The relation between the formerly introduced quantities of the particle 
is given by
\beqn
p^i &=& h^{i}_{\;\;\nu}({\bf k}) k^\nu \quad, \label{raise} \\
g_{\nu\kappa}({\bf k}) &=& \eta_{ij} h^{i}_{\;\;\nu}({\bf k}) h^{j}_{\;\;\kappa}({\bf k})\label{raise1}   \quad,
\eeqn 
where $\eta$ is the Minkowski metric. The dispersion relation reads simply $\eta_{ij} p^i p^j = 0$, or, 
more intuitively
\beqn
k_\nu g^{\nu \kappa}({\bf k}) k_\kappa = 0 \quad. \label{mdr}
\eeqn 
We will in the following refer to the dispersion relation being a modified dispersion relation ({\sc MDR}) if 
\beqn
\eta^{\kappa\nu} k_\kappa k_\nu \neq 0 \quad.
\eeqn
Note, that this need not necessarily be the case for all equations of the form (\ref{mdr}). E.g. when the
energy dependent metric is conformally flat and of the form $g^{\kappa \nu} = a(k) \eta^{\kappa \nu}$ with some scaling function $a$, then the dispersion relation (\ref{mdr}) implies the standard dispersion relation.

We can write the relation in the general form $p^i = G^i (k)$ with
\beqn
G^i(k) = \delta^{i}_{\;\; \nu} k^\nu + \sum_{l=1}^\infty \frac{{A_{(2l+1)}}^{i \nu_1 \nu_2 ... \nu_{2l+1}}}{m_{\rm p}^{2l}} 
k_{\nu_1}k_{\nu_2}...k_{\nu_{2l+1}}  \nonumber 
\eeqn 
where it is taken into account that $p$ is odd in $k$. $A$ is a rank-$2l+1$-tensor with dimensionless
coefficients that are constant with respect to space-time coordinates. 
Here, it was assumed that $m_{\rm p}$ sets the scale for the higher order terms.

Under quantization, the local quantity $k$ will be translated into a partial
derivate. One now wants to proceed from a single-$k$ mode 
\beqn
v_k \sim {\rm e}^{{\rm i} k_\nu x^\nu}
\eeqn 
to a field and to the operator ${\hat k}_\nu = -{\rm i}\partial_\nu$. The corresponding
momentum-operator $\hat p$ should have the property
\beqn
\hat p^i v_k = p^i v_k = G^i(k) v_k \quad,
\eeqn
which is fulfilled by
\beqn
\hat p^i = G^i(- {\rm i} \partial) \quad,
\eeqn
since every derivation results in just another factor $k$. It is therefore convenient to define the 
higher order operator 
\beqn
\delta^i = {\rm i} G^i (- {\rm i \partial}) \quad. \label{delta}
\eeqn
Since $G$ is even in $k$, this operator's expansion 
has only real coefficients that are up to signs those of $G^i$. 
Note that $\delta^i$ commutes with $\partial_\kappa$. A theory of this structure will usually
involve higher order derivatives in the spacelike as well as in the timelike coordinates that
require initial conditions. One thus expects the theory to have a rather complicated 
canonical structure, and to display inherently
non-local features. In particular the equal time commutation relations will be examined in section \ref{soc}.

From the above one can further define the operator $\widetilde{\Box}$  which generates 
the wave-function that corresponds to the {\sc MDR} Eq.(\ref{mdr})
\beqn
\widetilde{\Box} = g^{\mu\nu}(\partial_\alpha ) \partial_\mu \partial_\nu = \eta_{ij} \delta^i \delta^j \quad.\label{dal}
\eeqn
This modified D'Alembert operator plays the role of the propagator in the quantized theory.  
Normalized solutions to the wave-equation Eq. (\ref{eom}) can be found in the set of modes
\beqn
v_p(x) = \frac{1}{\sqrt{(2 \pi)^3 2 E}} {\exp} \left( {\rm i} k_\nu x^\nu \right) \label{pmode}
\eeqn
which solve the equation of motion
when $p$ fulfills the usual dispersion relation, or $k$ fulfills the {\sc MDR}, respectively\footnote{This
fixes the 'convenience factor' of \cite{Magueijo:2006qd} to just $E(\omega)$.}.  
Alternatively, one can
consider an expansion in $k$-space with $v_k = \sqrt{E/\omega} v_p$.
The solutions Eq.(\ref{pmode}) form an orthonormal set with respect to the new derivative
\beqn
\int {\rm d}^3 x~ v^*_p(x) \stackrel{\leftrightarrow}{\delta^0} v_{p'}(x) &=& \delta(k-k') \nonumber\\
&=& \delta(p-p') 
\bigg| \frac{\partial G}{\partial k } \bigg|~.
\eeqn
It is important to note that this complete set of orthonormal eigenfunctions of the 
momentum operator in the coordinate representation
are not also a complete set of eigenfunctions of the coordinate operator in the momentum 
representation, as it usually is the case. In $k$-space, the modes are normalized with respect to the
usual scalar product. Both descriptions are equivalent. The use of which is more suitable depends
on the quantity one wants to investigate. In case the standard momentum is $p$, it is more appropriate
to express everything in $p$-space. In the standard {\sc DSR}, one would instead want to formulate everything in
the modified quantity $k$.

The field expansion in terms of the set of solutions reads
\beqn
\phi(x) = \int {\rm d}^3 p  
\bigg| \frac{\partial F}{\partial p } \bigg|  \left[ v_p(x) a_p +v^*_p(x) a^\dag_p \right] \quad, \label{exp}
\eeqn
which yields the operators through forming the scalar product
\beqn
a_p &=& \int {\rm d}x^3 \left[(2\pi)^2 2 E\right]^{1/2} v^*_p(x) \stackrel{\leftrightarrow}{\delta^0} \phi(x) \\
a^\dag_p &=& \int {\rm d}x^3 \left[ (2\pi)^2 2 E \right]^{1/2} \phi(x) \stackrel{\leftrightarrow}{\delta^0} v_p(x)  \quad.
\eeqn
These fulfill the  commutation relation \cite{Hossenfelder:2003jz}
\beqn
[a_p,a^\dag_{p'}] = \delta(p-p') \bigg| \frac{\partial G}{\partial k}\bigg| \label{comm}\quad.
\eeqn

It is convenient to use the higher order operator $\delta^i$ in the setup of a field
theory, instead if having to deal with an explicit infinite sum.
Note, that this sum actually has to be infinite when the relation $p^i = G^i(k)$ has an 
asymptotic limit as one needs for an UV-regulator. Such asymptotic behavior can never be achieved with a finite power-series.

For the following analysis it is important to note that the higher order operator $\delta^i$ fulfills the property 
\beqn
\phi_i \left( \delta^i \psi \right) = - \left(\delta^i \phi_i \right) \psi + \mbox{total divergence}, \label{tbp}
\eeqn
which has been derived in \cite{Hossenfelder:2006cw}.
As an simple example we will work with a massless scalar field. The action for the scalar field \footnote{For a discussion of the Dirac-equation, gauge fields, and applications see e.g. 
\cite{Hossenfelder:2003jz,Hossenfelder:2004up}.} takes the form
\beqn
S &=& \int {\rm d}^4 x \sqrt{g }{\cal L} \quad,
\eeqn
with
\beqn
{\mathcal L} &=& \left( \partial_\nu \phi \right) \left( g^{\nu\kappa} \partial_\kappa \phi\right) \quad.
\eeqn
Using Eq.(\ref{tbp}), one then derives the equations from the usual variational principle to the correct form
\beqn
g^{\nu\kappa}\partial_\kappa \partial_\nu \phi = 0\quad. \label{eom}
\eeqn
The calculus with the higher order operator $\delta^i$ effectively summarizes the explicit 
dealing with the infinite series. These higher order derivative theories have been examined in \cite{Magueijo:2006qd}, where also the conserved
Noether currents have been derived and an explicit expression for the energy-momentum tensor can be found. 
For our purposes it is sufficient to note that
the Noether current is a bilinear form in the fields derivatives. If one inserts  the field expansion, integrates
it over space and takes the vacuum expectation value, one obtains (after normal ordering) the conserved quantity
\beqn
q^\nu = \int d^3 x \langle 0 |:T^{\nu}_{\;\; 0}:| 0 \rangle \quad, \label{tote}
\eeqn
which fulfils $\partial_\nu q^\nu =0$, and whose $0$-component can be identified as the total energy ${\mathcal E}$. If one inserts a superposition of  
two plane waves with ${\bf k}_1$ and ${\bf k}_2$, one 
finds that it is additive, since the mixing terms in the bilinear form do not contribute when the volume integration is performed. With this result from \cite{Magueijo:2006qd} the soccer-ball problem is absent. 
Due to the standard additivity, this expression for 
the total energy reduces to the usual 
expression already when the energy of each constituent is $\ll m_{\rm p}$.
In this case however, one has not only solved the multi-particle problem, but also removed the threshold modifications. In fact,
this result in incompatible with the {\sc DSR} interpretation in which the physically relevant 
and conserved quantity is obtained through a modified addition law.
 
As we had noticed before, this conserved quantity can then no longer be subject to the {\sc DSR} transformation,
the reason for which we can now identify. The above examination shows us very nicely where the problem stems from. 
It arises from the fact that the relation between the usual and the deformed quantity $h$ is a function only 
of the one mode it acts on, and so is the metric. If we apply it to the field's expansion, 
each term under the integral acquires a different transformation law, and we are back to the 
problem (\ref{notra}). 

In General Relativity however, the metric is a function not of the energy of a single mode, but of 
the energy-momentum tensor of the whole quantum field. In a theory of quantum gravity, the metric $g$
would become an operator, and the action would be a functional of $g$ coupled to the quantum field $\phi$. When
applying the variational principle, both are treated as independent variables. Variation with respect
to the field $\phi$ results in the field's equations of motion; variation with respect to the metric 
should result in a quantum version of Einstein's field equations. From dimensional
arguments, and to recover the classical limit, the source term in the latter equations should be the field's
density, and not a global charge. 

In lack of the full theory of quantum gravity, the here 
investigated approach can be understood as an educated guess for the arising metric. Instead of deriving it,
we required it to reproduce the existence of a minimal length which captures one of the best 
known, and most widely examined, properties of gravitational effects in the Planckian regime. This
metric then can be inserted in the field equations for $\phi$ which makes them non-linear. Nevertheless,
the conjectured metric operator should be a function of the field's densities, and instead
of it being a function of $k$ only, it should be of the form $g_{\mu \nu}( \partial \phi \partial \phi)$. Moreover,
it follows from this that the relation between the momentum $p$ and the wave vector $k$ of a single mode therefore does {\sl not}
only depend on the mode's properties, but on the energy density of the whole field and Eq. (\ref{raise}) should
correctly read
\beqn
p^i = h^{i}_{\;\; \nu}(\partial \phi \partial \phi) k^\nu \quad. \label{raise2}
\eeqn
In contrast to the single particles that were considered for the construction of the original {\sc DSR} 
transformations, plane waves do overlap each other. The transformations acting on one wave will therefore
be sensitive to the energy content of the other waves, all of which taken together 
determine the structure of the momentum space bundle over the space-time. Up to dimensional factors, 
the standard {\sc DSR} approach remains applicable for a single mode, in which case the energy density is proportional 
to the mode's frequency.

\section{The soccer ball problem}
\label{soc}

One of the truly surprising features of {\sc DSR} is that a particle
of a very tiny mass compared to the Planck scale can perceive {\sc DSR} effects that are 
argued to be caused by quantum gravity. Naively, one would expect quantum gravitational 
effects to become important only when the curvature of the background is non-negligible. 
This is usually not the case for particles we observe.

This reflects in the above finding that the relation between the quantity with the standard properties ${\bf p}$ and
the modified one ${\bf k}$ should be a function of the energy-momentum density rather than the total energy. 
One should also keep in mind that under quantization, 
modifications of the type ${\bf k}=F({\bf p})$ lead to a modified commutation relation \cite{Hossenfelder:2005ed,Kempf:1994su,ml1,ml2,ml3}
which results in a generalized uncertainty relation. This generalized uncertainty makes it 
impossible to localize a particle to better precision than a Planck
length, which is what one would expect. 

One reproduces the equivalent of the generalized uncertainty
for a quantum field theory by considering the commutator of the field and its conjugated variable.
With the help of the previously defined higher order operator $\delta^\nu$, one can define
a conjugated momentum of the field to
\beqn
\pi^\nu = \delta^\nu \phi = \frac{\partial {\mathcal L}}{\partial (\partial_\nu \phi)}\quad, \label{pi}
\eeqn
with the identification $\pi^0 \equiv \pi$.
  
From $\pi=\delta^0 \phi(y)$ with use of the field expansion Eq.(\ref{exp}) one then finds
in the usual way
 
\beqn
\left[\phi(x), \pi^0(y) \right]  &=& {\rm i}  \int {\rm d}^3 p \bigg| \frac{\partial F}{\partial p }  \bigg|    
\int {\rm d}^3 p' 
\bigg| \frac{\partial F'}{\partial p' } \bigg| \times \nonumber \\
&&  2 E'  v_p(x) v^*_{p'}(y) \left[ a_p,a_{p'}^\dag\right] \quad.
\eeqn
Using Eq.(\ref{comm}) this reduces to 
\beqn
\left[\phi(x), \pi(y) \right]   
&=& {\rm i} \int {\rm d}^3 p \bigg| \frac{\partial F}{\partial p }  \bigg| 2 E v_p(x) v^*_{p}(y) \nonumber \\
&=& {\rm i} \int \frac{{\rm d}^3 p}{(2 \pi)^3} \bigg| \frac{\partial F}{\partial p }  
\bigg| {\rm e}^{ {\rm i}k (x-y)}
\quad.
\eeqn

A specifically useful relation from \cite{Hossenfelder:2004up} for $k(p)$ is 
\beqn
k_{\mu}(p) &=& \hat{e}_{\mu} \int_0^{p} e^{{\displaystyle{-\epsilon p'^2}}} {\rm d} p'\label{model} \quad,\label{example}
\eeqn
where $\hat{e}_{\mu}$ is the unit vector in $\mu$ direction, $p^2=\vec{p}\cdot\vec{p}$ and
$\epsilon=l_{\rm{p}}^2 \pi / 4 $ (the factor $\pi/4$ is included to assure, that the 
limiting value is indeed $1/l_{\rm{p}}$).
Using this example one finds
\beqn
\left[\phi(x), \pi(y) \right]   
&=& {\rm i} \int \frac{{\rm d}^3 p}{(2 \pi)^3}  {\rm e}^{ {\rm i}k (x-y) - \varepsilon p^2 }
\quad.
\eeqn
One sees that a non-trivial dispersion relation with a lower bound on the wave-length 
therefore implies a locality bound similar to that proposed in \cite{Giddings:2001pt,Giddings:2004ud}
\beqn
\left[\phi(x), \pi(y) \right]   \neq {\rm i} \delta(x-y)   
\quad,
\eeqn
which is due to the non-trivial volume element in momentum space.  
Rewriting the expression into $k$-space, one realizes that
this arises through the finite boundaries. Such a modification will become 
important, when $x-y \sim l_{\rm p}$. 

%In this approach we have implicitly assumed that the quantization procedure builds up
%on the commutation relations of the creation and annihilation operators (in momentum space), which was
%also used in \cite{Magueijo:2006qd}.

It therefore seems inappropriate to consider an integrated
quantity that can not be localized to a point particle by using superpositions, since this is
disabled by the very postulate of a minimal length. Instead, one should consider the local density of the field, and
impose a bound on it. This is also a more appropriate choice simply because we want to construct a field 
theory for {\sc DSR}. 

In the standard {\sc DSR} approach there is no dependence on 
the volume inside which we localize a mode with a given energy. We can use box modes and shrink
the box as small as possible, that is as small as a Planck-volume. This does not reflect in any way in
the transformation properties of the modes. If one thinks in terms of total energy, then it is not even clear 
in which limit an un-deformed Special Relativity has to be recovered. The limit of a total energy ${\mathcal E}$ 
very small with respect to the Planck-mass, ${\mathcal E} \ll m_{\rm p}$ (single particle), as well as 
very large  $m_{\rm p} \ll {\mathcal E}$ (multi-particle) need to reduce to the standard transformation behavior, 
since we have observations in both cases that show no deviation from Special Relativity. Instead, the limit 
that one would like to take is that of a 
small energy density with respect to the Planck scale ${\mathcal E}/{\mathrm{Volume}} \ll {m_{\rm p}}/l_{\rm p}^3$.
This means however, that the whole formalism of {\sc DSR} needs to be sensitive to the volume in
which we localize the energy. 

Furthermore, let us recall what we found earlier that {\sc DSR} can be understood as a theory with
a curved momentum space. For a single particle, we were just concerned with the particle's configuration
space over the particle's world line. As long as the particles are separated from each other, it is
conceivable to treat their momentum spaces as independent. In the case of superpositions of modes however, the
properties of the momentum spaces over the space-time in which the field extends should depend on all
of the modes that contribute to the field's composition. 

We are therefore lead to the conclusion that the quantity to be bounded in {\sc DSR} should not be the energy of a
particle,  but rather the energy density of a matter field. 

One sees now easily that the soccer-ball problem arises from the fact that the {\sc DSR}-formalism 
does not forbid us to consider multi-particle systems inside a region of spacetime possibly as 
small as $l_{\rm p}^3$, but with an unlimited total energy of the particles. However, when we 
go from a microscopic system to a macroscopic system in a sequence like quark $\to$ proton $\to$ nucleus $\to$ atom $\to$ soccer-ball, then 
the number of constituents grows, and so does the total energy, but the energy density usually drops. Consequently, one would expect 
gravitational effects to become less important. In the usual {\sc DSR} approach however, it is 
possible to place an arbitrary amount of particles arbitrarily close to each other. This is not only 
inconsistent with the ansatz itself to understand {\sc DSR} as a an effective quantum gravitational description 
(since the system would inevitably undergo gravitational collapse), but it is also in conflict with 
every day experience. 

It is also important to note that for a quantized theory the number of constituents of an object is a very ill defined quantity, due to virtual particle content, and would better be avoided.

Again, we are lead to the conclusion that the quantity to consider should be the field's energy-momentum density 
rather than the four-momentum of a particle. Or, to construct a four-vector, the projection of 
the energy-momentum tensor on 
an observer's four velocity $u_\kappa$:
\beqn
J^\nu:= T^{\kappa\nu} u_\kappa \leq m_{\rm p}^4 \quad.
\eeqn
In the restframe this is just ${\bf J} = (\rho,0,0,0)$. Like for the momentum, one derives easily a deformed 
transformation that respects this behavior by just replacing $\omega \to J^0, k \to J$, 
and $m_{\rm p} \to m_{\rm p}^4$ in Eq. (\ref{trafok}). 

Now what happened to the soccer-ball problem? Well, a system's energy density is not an extensive quantity. 
Thus, there is no problem with an addition law. In fact, we notice immediately that the 
deformations and modifications vanish as one would expect for energy densities $\rho \ll m_{\rm p}^4$, and 
soccer-balls do exist and transform like we are used to from soccer-balls. 

If one considers the addition of energies of single modes, one arrives via Noether's theorem at the above derived 
quantity Eq.(\ref{tote}) \footnote{Note that the metric is an independent variable and not varied with respect to $\phi$. 
The dependence of $g$ on $\phi$ arises from inserting a conjectured solution to the field equations for $g$.}. To
be in accordance with the {\sc DSR} interpretation, we again lower the index and arise at the addition law 
Eq.(\ref{c2}) for the quantity with the deformed transformation behavior. The previously encountered problem that
this quantity does not transform covariantly (\ref{notra}) is absent because the transformation $\widetilde L$ now is not 
only a function of the mode it acts on, but a function of both modes' wave-vectors since both enter ${\bf J}$.
Therefore, the transformations for both modes are equal. In this way, the total energy of a system can become
arbitrarily large, but its transformation properties approach the Special Relativistic limit for small densities. 
 
One finds the common {\sc DSR} prescription for a particle with ${\bf k}= (\omega ,k)$ if one localizes it as good 
as maximally possible while still respecting the generalized uncertainty, that is one 
sets ${\bf J} m_{\rm p}^3  = {\bf k}$.

\section{Predictions Revisited}
\label{pred}

These were the good news. Now to the bad news. The time of flight analysis for photons
with different energies from $\gamma$-ray bursts has been proposed as a possible test for
{\sc DSR}. If the speed of light is energy dependent, one finds a
possible time delay $\Delta T$ of \cite{Scargle:2006kr,Amelino-Camelia:2003bt}
\beqn
\Delta T \sim T \frac{E}{m_{\rm p}}\quad,
\eeqn
where $T$ is the duration of travel, which for a distance as large as a Gpc can be up to $\sim 10^{17}$~s. Even 
though for typical energies like $E\sim$~GeV, the ratio to the Planck mass is tiny
 $E/m_{\rm p}\sim 10^{-19}$, the long distance traveled results in $\Delta T \sim 10^{-2}$s. This time 
delay is comparable
to the typical duration of the burst, and thus potentially measurable.

However, having come to the conclusion that not the energy of a single particle is the relevant quantity
but rather its energy density that curves the space it propagates in, let us repeat this analysis.  
The typical peak energy of a $\gamma$-ray burst is $\sim 100$~keV, but let us consider one of highest 
peak energy $\sim$~GeV, which will have a typical localization of $\sim$~fm, and an energy density of
roughly $\rho \sim 10^{-76} m_{\rm p}/l_{\rm p}^3$.  Thus, the effect is about $57$ orders of magnitude smaller than predicted\footnote{The brightness of a $\gamma$-ray burst corresponds to an energy of the total mass of the sun 
$\sim 10^{54}$~TeV released in only a few seconds. Even under this extreme conditions in the initial moments
of the burst, the background curvature is far from being subject to quantum gravity.}. 

It should be pointed out that this conclusion does not apply for theories with a modified dispersion 
relation that explicitly break Lorentz-invariance. All the here made investigations are based on the 
assumption of observer-independence without a preferred frame. 
 
\section{Discussion and Conclusion}
\label{conc}

In the previous sections we have seen that the transfer of the single particle {\sc DSR} prescription to a
field theory needs to be formulated in the field's densities rather than in integrated quantities of total
four momentum. The
dependence on the volume inside which energy is accumulated is necessary to recover the standard
transformation behavior in the limit when the density is small compared to the Planck scale, and
quantum gravitational effects can safely be neglected. It has also been previously pointed 
out in \cite{Girelli:2006ez} that the soccer-ball problem might be due to the use of quantities with 
inappropriate dimensionality. Another related approach, suggested in \cite{Magueijo:2006qd}, is that the relevant quantity could scale with the number of constituents. 

We have seen that the soccer-ball problem is naturally absent if one assumes a deformation of
Special Relativity that saturates with the energy density approaching the Planckian limit. The total
energy adds linearly, but its transformation is deformed as a function of the energy density. The
total energy of macroscopic systems thus can exceed the Planck mass, while the dropping 
energy density assures the recovery of Special Relativity. As a consequence, predictions for the
measurement of an energy dependent speed of light with $\gamma$-ray bursts are rendered unobservable, since
the scale for the effect is set by the typical energy density rather than the total energy of the photons.

Now why am I writing such a depressing paper? The reason is that despite the 
excitement that {\sc DSR} has understandably caused, one should not neglect 
the demand for consistency. A model that is
claimed to potentially describe nature must reproduce the known and well 
established theories in the range that we have confirmed them with observations. It is not
difficult to make exciting predictions if one weakens this requirement. Even though I
find the possibility to experimentally test quantum gravity extremely fascinating, 
one should carefully investigate the known problems of the approach as to whether
they are fatal.  

Deformed Special Relativity, in the interpretation as commonly used, is not able
to reproduce the Standard Model of particle physics because multi-particle states 
can not be described. For the same reason, it is not possible to reproduce the usual
transformation laws of Special Relativity for macroscopic objects.  

The here presented analysis does not aim to provide a complete quantum field theory with {\sc DSR} that 
incorporates the suggested framework, but it presents a starting point for further investigations. 
I am summarizing the difficulties with the common approach here not because I like to tell 
depressing stories, but because I think that it is indeed possible
to formulate a quantum field theory with {\sc DSR}, that does not suffer from the above
mentioned problems. This theory might be less exciting but also less speculative. 
 
\begin{acknowledgments}
I thank Stefan Hofmann, Tomasz Konopka and Lee Smolin for helpful discussions. 
Research at Perimeter Institute for Theoretical Physics is supported in
part by the Government of Canada through {\sc NSERC} and by the Province of Ontario through {\sc MRI.}

\end{acknowledgments}

\end{document}